\documentstyle[preprint,epsfig,aps,amssymb]{revtex}
\begin{document}
\draft
\narrowtext
\tighten
\title{P, T-Violating Electron-Nucleon Interactions in the\\
R-Parity Violating Minimal Supersymmetric Standard Model}
\author{Peter Herczeg}
\address{Los Alamos National Laboratory, Theoretical Division\\
Los Alamos, NM  87545}
\date{December 7, 1999}
\preprint{LA-UR-99-6562}
\maketitle
\begin{abstract}
We show that the present experimental limits on electron-nucleon interactions
that violate both parity and time reversal invariance provide new stringent
bounds on the imaginary parts of some of the products of the R-parity violating
coupling constants in the R-parity violating Minimal Supersymmetric Standard
Model.
\end{abstract}
\vspace{.5 in}
\section{Introduction}

In the Minimal Supersymmetric Standard Model (MSSM) \cite{1}, unlike
in the Standard Model (SM) \cite{2}, the conservation of lepton number ($L$)
and of baryon number ($B$) is not automatic \cite{3,4}.  In particular, the
superpotential can contain renormalizable and gauge invariant $L-$ and
$B-$violating terms.  The general forms of these are \cite{4}
\begin{equation}
W_{\not{L}} = \frac{1}{2} \lambda_{ijk} L_i L_j E^c_k +
\lambda^\prime_{ijk} L_i Q_j D^c_k + \mu_i L_i H_u\; ,
\end{equation}
\begin{equation}
W_{\not{B}} = \frac{1}{2} \lambda^{\prime \prime}_{ijk} U^c_i D^c_j
D^c_k\; ,
\end{equation}
where $i, j, k = 1, 2, 3$ are family indices, and summations over $i,
j, k$ are implied.  In Eqs. (1) and (2) $L_i, Q_i$ are the
S$U(2)$-doublet lepton and quark superfields, $E^c_i$, $U^c_i$,
$D^c_i$ are the $SU(2)$-singlet charged lepton, and up- and down-type
quark superfields; $H_u$ is the Higgs superfield which generates the
masses of the up-type quarks.  The constants $\lambda_{ijk}$ are
antisymmetric under the interchange $i \leftrightarrow j$, and
$\lambda^{\prime \prime}_{ijk}$ is antisymmetric under $j
\leftrightarrow k$.

The  couplings in $W_{\not{L}}$  and $W_{\not{B}}$  violate invariance
under $R$-parity ($R  = (-1)^{3B+L+2s}$, where $s$ is  the spin of the
particle; thus $R = +1$ for the  particles of the SM, and $R = -1$ for
their       superpartners)      \cite{5}.       If       both      the
$\lambda^{\prime}_{ijk}$-term       and      the      $\lambda^{\prime
\prime}_{ijk}$-term is present, some of  the products would have to be
extremely  small (for example,  $|\lambda^\prime_{11k} \lambda^{\prime
\prime}_{11k}| \lesssim 10^{-22}$   for   $k   =  2,   3$   and
$m_{\tilde{d}_{kR}} =  100 \ GeV$)  to prevent too rapid  proton decay
\cite{6}.   One  way  to  deal  with  this  problem  is  to  postulate
$R$-parity  invariance.  This would  eliminate both  $W_{\not{L}}$ and
$W_{\not{B}}$ \cite{4}.  Another possibility is that $B$ is conserved,
but the $L$-violating terms are present.  This scenario is obtained by
demanding invariance under ``baryon parity'' (under baryon parity $Q_i
\rightarrow  -Q_i,$  $U^c_i  \rightarrow -U^c_i$,  $D^c_i  \rightarrow
-D^c_i$,  and $L_i$,  $E^c_i$, $H_u$,  $H_d$ remain  unchanged) \cite{4,7}.
The  model  we shall  consider  in  the  following is  the  $R$-parity
violating  MSSM (R\hspace{-,12in}/MSSM),  defined as  the MSSM  with
$W_{\not{L}}$ included in the superpotential \cite{8}.

The presence of $R$-parity violating couplings has rich
phenomenological implications.  If $R$-parity is violated, the
production of single supersymmetric particles becomes possible, and
the lightest supersymmetric particle is no longer stable.  The main
source of constraints on the $R$-parity violating coupling constants
is experimental data on processes with the SM particles, to which the
$R$-parity violating couplings can contribute through the exchange of
single squarks or sleptons.  The present status of these bounds is
given in the reviews in Ref. \cite{9}.  Most of the upper limits on
the individual coupling constants are of the order of $10^{-2} -
10^{-1}$ for squark and slepton masses of $100$ GeV.  There are also
more stringent bounds, including some on products of two coupling
constants, coming mainly from processes forbidden in the SM.

The $R$-parity violating coupling constants can be complex, and thus
represent new sources of CP-violation.  Stringent constraints come
from the experimental values of $\epsilon$ and
$\epsilon^\prime/\epsilon$ \cite{10}.  The effects of CP-violating
$R$-parity violating interactions have also been considered in
semileptonic K-decays \cite{11}, B-decays \cite{12}, semi-inclusive
decays of heavy quarks \cite{13}, leptonic collider processes
\cite{14}, and in lepton-pair production in $\overline{p}p$ reactions
\cite{15}.

In this paper we show that the present experimental limits on
$P$, $T$-violating e-N interactions (electron-nucleon interactions that
violate both parity and time reversal invariance) set stringent bounds
on the imaginary parts of some of the products $\lambda^*_{ijk}
\lambda^\prime_{\ell mn}$.  In the next section we analyze the $P$,
$T$-violating e-N interactions arising from the $R$-parity violating
couplings.  In Section III we derive the bounds on the imaginary parts
of products of the coupling constants involved, and consider the
constraints on them from other data.  In Section IV we summarize our
conclusions.

\section{$P$, $T$-violating \lowercase{e}-N interactions in the
R\hspace{-.12in}/MSSM}

The general form of $P$, $T$-violating e-N interactions, including
non-derivative couplings only, is given by \cite{16}
\begin{eqnarray}
H_{P,T} & = & \sum_{a=p,n} \frac{G_F}{\sqrt{2}} [C_{Sa} \overline{e}i\gamma_5e
\overline{a}a + C_{Pa} \overline{e}e \overline{a}i\gamma_5a \nonumber \\
& & \mbox{} + C_{Ta} \frac{1}{2} i\epsilon_{\alpha \beta \gamma 
\delta} \overline{e}
\sigma^{\alpha \beta} e \overline{a} \sigma^{\gamma \delta} a ] \; ,
\end{eqnarray}
where $a = p,n$ ($p$ = proton, $n$ = neutron) and $C_{Sa}$, $C_{Pa}$
and $C_{Ta}$ are real constants \cite{17}.

Stringent limits on $C_{Sa}$, $C_{Pa}$ and $C_{Ta}$ \cite{18} follow from
experimental results on the electric dipole moments of the $^{133}Cs$
\cite{19}, $^{205}T\ell$ \cite{20}, $^{129}Xe$ \cite{21} and 
$^{199}Hg$ \cite{22} atoms,
and on the P,T-violating spin-flip parameter $\nu$ of the $T\ell F$
molecule \cite{23}.  The best of these are \cite{16}
\begin{equation}
\left| 0.4 C_{Sp} + 0.6 C_{Sn} \right| < 3.4 \times 10^{-7}\;,
\end{equation}
\begin{equation}
\left| C_{Pn} \right| < 1.4 \times 10^{-5}\; ,
\end{equation}
\begin{equation}
\left| 0.75 C_{Pp} + 0.25 C_{Pn} \right| < 3 \times 10^{-4} \; ,
\end{equation}
\begin{equation}
\left| C_{Tn} \right| < 4 \times 10^{-8} \; ,
\end{equation}
\begin{equation}
|0.75 C_{Tp} + 0.25 C_{Tn}| < 4.5 \times 10^{-7} \; ,
\end{equation}
The limit (4) has been deduced from the experimental bound on $d(T\ell)$;
the limits (5) and (7) come from $d(Hg)$, and the limits (6) and (8) from
$\nu(T\ell F)$.

In the SM the constants $C_{Sa}$, $C_{Pa}$ and $C_{Ta}$ are very
small: the Kobayashi-Maskawa phase contributes at the level of
$10^{-16}$, and the contribution of the $\theta$-term is of the order
of $10^{-11} - 10^{-10}$ \cite{24}.  They can be much larger, however, in
some extensions of the SM.  The $P$, $T$-violating e-N interactions have
been studied for general electron-quark interactions \cite{25}, in
multi-Higgs models \cite{24,26,27}, models with leptoquarks 
\cite{24,27,28}, and
in the $R$-parity conserving MSSM \cite{29}.  In multi-Higgs and leptoquark
models they can be as strong as allowed by the present experimental
limits.  In the $R$-parity conserving MSSM the constants $C_{Sa}$ (which
are the dominant ones \cite{29}) are smaller by several orders of magnitude
than the present limit on $C_{Sa}$ [Eq. (4)].  This is implied by the
limit on the pertinent CP-violating phase, obtained from the limit on
the $P$, $T$-violating nucleon-nucleon interactions set by the
experimental limit on $d(Hg)$ \cite{30}.

We shall consider now the $P$, $T$-violating e-N interactions in the
R\hspace{-.12in}/MSSM.  Relative to the $R$-parity conserving MSSM, in this
model there are additional contributions to $P$, $T$-violating e-N
interactions, originating from the $\lambda_{ijk}$ and
$\lambda^\prime_{ijk}$ couplings in Eq. (1), which appear already at
the tree level.  The $\lambda_{ijk}$ and $\lambda^\prime_{ijk}$
couplings in (1) in terms of the components of the superfields are
given by \cite{31}
\begin{eqnarray}
{\mathcal{L}} & = & \lambda_{ijk} [\tilde{\nu}_{iL}
\overline{e}_{kR} e_{jL} + \tilde{e}_{jL} \overline{e}_{kR} \nu_{iL}
+ \tilde{e}_{kR}^* \overline{\nu^c}_{iL} e_{jL} \nonumber \\
& & \mbox{} - \tilde{\nu}_{jL} \overline{e}_{kR} e_{iL} - \tilde{e}_{iL}
\overline{e}_{kR} \nu_{jL} - \tilde{e}_{kR}^* \overline{\nu^c}_{jL}
e_{iL} ]  \nonumber \\
& & \mbox{} + \lambda^\prime_{ijk} [\tilde{\nu}_{iL} \overline{d}_{kR} d_{jL} +
\tilde{d}_{jL} \overline{d}_{kR} \nu_{iL} + \tilde{d}_{kR}^*
\overline{\nu^c}_{iL} d_{jL} \nonumber \\
& & \mbox{} - \tilde{e}_{iL} \overline{d}_{kR} u_{jL} - \tilde{u}_{jL}
\overline{d}_{kR} e_{iL} - \tilde{d}_{kR}^* \overline{e^c}_{iL} u_{jL} ] +
H. c.
\end{eqnarray}
In Eq. (9) only the $\lambda_{ijk}$ with $i < j$ are nonvanishing, since the
relation $\lambda_{jik} = -\lambda_{ijk}$ that holds for the coupling
constants in the first term in Eq. (1) has already been used.

Contributions to electron-quark interactions can come from the
combination of either two $\lambda^\prime_{ijk}$ terms, or from a
combination of a $\lambda_{ijk}$ and a $\lambda^\prime_{ijk}$ term.
It is easy to see that in the former case the electron-quark
interaction has no P, T-violating component.  The reason is that the
only terms in (9) involving the electron field are
$\tilde{u}_{jL}\overline{d}_{kR}e_{iL}$ and $\tilde{d}^*_{kR}
\overline{e^c}_{iL} u_{jL}$, and therefore their contribution to the
electron-quark interactions must be proportional to
$|\lambda^\prime_{ijk}|^2$, which is insensitive to CP-violating
phases.  $P$, $T$-violating contributions do arise however from
combinations of a $\lambda_{ijk}$ and a $\lambda^\prime_{ijk}$ term.
We find that there are two such contributions for each down-type
quark: one mediated by $\tilde{\nu}_{\mu L}$, and one by
$\tilde{\nu}_{\tau L}$ (see Fig. 1).  The corresponding effective
Hamiltonian is given by
\begin{equation}
H = \sum_{k=1,2,3} \sum_{j=2,3} \frac{\lambda^*_{1j1} \lambda^\prime_{jkk}}{4
m^2_{\tilde{\nu}_j}} \overline{e}(1 +  \gamma_{5})e \overline{d}_k (1 -
\gamma_5)d_k + H.c.
\end{equation}
The $P$, $T$-violating component of (10) is
\begin{equation}
H_{P,T} = \sum_{k=1,2,3} \sum_{j=2,3} \frac{ {\rm Im} \left(
\lambda^*_{1j1} \lambda^\prime_{jkk} \right)}{2 m^2_{\tilde{\nu}_j}}
(\overline{e} i\gamma_5 e \overline{d}_k d_k - \overline{e} e
\overline{d}_k i \gamma_5 d_k) \; .
\end{equation}
It follows that the constants in the e-N interaction (3) are given by
\begin{equation}
C_{Sa} = \sum_{k=1,2,3} \sum_{j=2,3} \frac{{\rm Im} \left(
\lambda^*_{1j1} \lambda^\prime_{jkk} \right)}{2 m^2_{\tilde{\nu}_j}}
\frac{\sqrt{2}}{G_F} f_a^{(d_k)} \; ,
\end{equation}
\begin{equation}
C_{Pa} = -\sum_{k=1,2,3} \sum_{j=2,3} \frac{{\rm Im}\left(
\lambda^*_{1j1} \lambda^\prime_{jkk} \right)}{2 m^2_{\tilde{\nu}_j}}
\frac{\sqrt{2}}{G_F} g_a^{(d_k)} \; ,
\end{equation}
\begin{equation}
C_{Ta} = 0 \; ,
\end{equation}
where $f_a^{(d_k)}$ and $g_a^{(d_k)}$ are defined by
\begin{equation}
\left\langle a \left| \overline{d}_k d_k \right| a \right\rangle =
f_a^{(d_k)} \overline{u}_a u_a \ \ \ \ \ \ \ \ \ (k = 1,2,3; \ a =
p,n) \; ,
\end{equation}
\begin{equation}
\left\langle a \left| \overline{d}_k i \gamma_5 d_k \right| a
\right\rangle = g_a^{(d_k)} \overline{u}_a i \gamma_5 u_a \ \ \ \ \ \
\ \ \ (k = 1,2,3; \ a = p,n) \; .
\end{equation}
An estimate of the matrix elements $\left\langle p \left| \overline{d}
d \right| p \right\rangle $ and $\left\langle p \left| \overline{s} s
\right| p \right\rangle$ can be obtained from the $\sigma$-term
(deduced from pion-nucleon scattering data) and, assuming octet type
$SU(3)$ breaking, from baryon mass splitting \cite{32}.  These yield \cite{33}
\begin{equation}
f_p^{(d)} \simeq 2.8 \; ,
\end{equation}
\begin{equation}
f_p^{(s)} \simeq 1.4 \; .
\end{equation}
One obtains also $f^{(u)}_p \simeq 3.5$ for the form factor in the
matrix element $\left\langle p \left| \overline{u} u \right| p
\right\rangle = f^{(u)}_p \overline{u}_p u_p$. Charge symmetry and
isospin invariance imply, respectively, $f_n^{(d)} = f_p^{(u)}$ and
$f_n^{(s)} = f_p^{(s)}$, so that
\begin{equation}
f_n^{(d)} \simeq 3.5 \; ,
\end{equation}
\begin{equation}
f_n^{(s)} \simeq 1.4 \; .
\end{equation}

The matrix elements $\left\langle a \left| \overline{d} i \gamma_5 d
\right| a \right\rangle$ and $\left\langle a \left| \overline{s} i
\gamma_5 s \right| a \right\rangle$ $(a = p,n)$ can be estimated using
the relation (see Ref. \cite{34})
\begin{equation}
g^{(q)}_p = \frac{M_p}{m_q} (\Delta q^\prime + \frac{\alpha_s}{2\pi}
\Delta g) \
\ \ \ \ \ \ \ (q = u,d,s) \; ,
\end{equation}
where $M_p$ is the proton mass, $\Delta q^\prime$ is the form factor
at zero momentum transfer in the proton matrix element of the axial
vector current $ (\left\langle p \left| \overline{q} \gamma_\lambda
\gamma_5 q \right| p \right\rangle$ = $\Delta q^\prime \overline{u}_p
\gamma_\lambda \gamma_5 u_p)$ and $\Delta g$ is defined by
\begin{equation}
\langle p | Tr G_{\mu \nu} \tilde{G}^{\mu \nu} | p \rangle = - 2 M_p \Delta g
\overline{u}_p i \gamma_5 u_p \; ,
\end{equation}
where $G_{\mu \nu}$ is the gluon field intensity and $\tilde{G}^{\mu
\nu} = \frac{1}{2} \epsilon^{\mu \nu \lambda \rho} G_{\lambda \rho}$.
With $\Delta u^\prime = 0.82$, $\Delta d^\prime = -0.44$, $\Delta
s^\prime = 0.11$ \cite{35}, deduced from data on polarized nucleon
structure functions, and the estimate $(\alpha_s / 2 \pi) \Delta g
\simeq - 0.16$ \cite{34}, we find
\begin{equation}
g^{(d)}_p \simeq 61 \; ,
\end{equation}
\begin{equation}
g^{(d)}_n  = g^{(u)}_p \simeq 121 \; ,
\end{equation}
\begin{equation}
g^{(s)}_n  = g^{(s)}_p \simeq -1.5 \; ,
\end{equation}
In Eqs. (24) and (25) we used again charge symmetry and isospin invariance.

There are no estimates available for $f_a^{(b)}$ and $g_a^{(b)}$.  We
shall use $f_a^{(b)}/f_a^{(c)} \simeq m_c/m_b$, $g_a^{(b)}/g_a^{(c)}
\simeq m_c/m_b$ (as expected on the basis of the heavy quark expansion
of the corresponding operators (see Ref. \cite{36})) and $f_a^{(c)} \simeq
0.04$, $g_a^{(c)} \simeq -0.1$ \cite{36}.  Then
\begin{equation}
f_a^{(b)} \simeq 10^{-2} \ \ \ \ \ \ \ \ \ \ \ (a = p,n) \; ,
\end{equation}
\begin{equation}
g_a^{(b)} \simeq 3 \times 10^{-2} \ \ \ \ \ \ \ (a = p,n) \; .
\end{equation}

\section{Bounds on the coupling constants}

We are now ready to consider the limits on the quantities
${\rm Im}(\lambda^*_{1j1} \lambda^\prime_{jkk})$.  In deriving these we
shall assume for each such product that it is the only one which may
have a significant size.  This assumption precludes additional
constraints to apply on a given ${\rm Im}(\lambda^*_{1j1}
\lambda^\prime_{jkk})$, and the possibility of cancellations among the
various contributions to $C_{Sa}$ and $C_{Pa}$.

\subsection{$\bbox{\lambda^*_{121}} \bbox{\lambda^\prime_{211}}$ and
$\bbox{\lambda^*_{131} \lambda^\prime_{311}}$}

The contribution of $\lambda^*_{121} \lambda^\prime_{211}$ to the
constants $C_{Sa}$ and $C_{Pa}$ is [see Eqs. (12) and (13)]
\begin{equation}
C_{Sa} = \frac{{\rm Im}(\lambda^*_{121}
\lambda^\prime_{211})}{2m^2_{\tilde{\nu}_\mu}}
\frac{\sqrt{2}}{G_F} f^{(d)}_a \ \ \ \ \ \ (a = p,n) \; ,
\end{equation}
\begin{equation}
C_{Pa} = \frac{{\rm Im}(\lambda^*_{121}
\lambda^\prime_{211})}{2m^2_{\tilde{\nu}_\mu}}
\frac{\sqrt{2}}{G_F} g^{(d)}_a \ \ \ \ \ \ (a = p,n) \; .
\end{equation}
We shall consider first the information from $d(T\ell)$.  We note that
although the ratio $g_a^{(d)}/f_a^{(d)}$ is large ($\sim 20$ and $\sim
40$ for $a=p$ and $a=n$, respectively), the contribution of the
$C_{Pa}$-interaction to $d(T\ell)$ can be neglected, since it is
supressed by about four orders of magnitude relative to the
contribution from $C_{Sa}$ (two orders of magnitude due to the absence
of enhancement by factors of $Z$ and $N$, and a further two orders of
magnitude due to the fact that the $C_{Pa}$-interaction arises only as
a correction to the nonrelativistic approximation).  Using for
$f^{(d)}_p$ and $f^{(d)}_n$ the values (17) and (19), we obtain from
the limit (4)
\begin{equation}
\left| {\rm Im}(\lambda^*_{121} \lambda^\prime_{211}) \right| \lesssim
1.7 \times 10^{-8} (m_{\tilde{\nu}_\mu}/100 \mbox{ GeV})^2 \; .
\end{equation}

With the value (24) for $g_n^{(d)}$ about the same limit ($\left| {\rm
Im}(\lambda^*_{121} \: \lambda^\prime_{211}) \right| \lesssim 1.9
\times 10^{-8}$ $(m_{\tilde{\nu}_\mu}/100$~GeV)$^2$) follows from
$d(Hg)$ [Eq. (5)].  In $d(Hg)$ the contribution of $C_{Pa}$ is smaller
than the contribution of $C_{Sa}$ only by an order of magnitude.  The
reason is that in diamagnetic atoms a $C_{Sa}$-interaction can
contribute to the electric dipole moment only with the participation
of the hyperfine interaction \cite{37}.  As a consequence, the limit
on $\left| {\rm Im}\left( \lambda^*_{121} \lambda^\prime_{211} \right)
\right|$ from the $C_{Sa}$-contribution is weaker then the one implied
by the $C_{Pa}$-contribution by about a factor of four.

For $\lambda^*_{131} \lambda^\prime_{311}$ one obtains in the same way
as for $\lambda^*_{121} \lambda^\prime_{211}$
\begin{equation}
\left| {\rm Im} \left( \lambda^*_{131} \lambda^\prime_{311} \right)
\right| \lesssim 1.7 \times 10^{-8} (m_{\tilde{\nu}_\tau}/ 100\mbox{
GeV})^2 \;.
\end{equation}

\subsection{$\bbox{\lambda^*_{121} \lambda^\prime_{222}}$ and
$\bbox{\lambda^*_{131} \lambda^\prime_{322}}$}

With the values (18) and (20) for $f_p^{(s)}$ and $f_n^{(s)}$ the
limit (4) from $d(T\ell)$ yields
\begin{equation}
\left| {\rm Im}\left( \lambda^*_{121} \lambda^\prime_{222}
\right)\right| \lesssim 4 \times 10^{-8} (m_{\tilde{\nu}_\mu}/ 100
\mbox{ GeV})^2 \; ,
\end{equation}
\begin{equation}
\left| {\rm Im} \left( \lambda^*_{131} \lambda^\prime_{322} \right)
\right| \lesssim 4 \times 10^{-8}
(m_{\tilde{\nu}_\tau}/ 100\mbox{ GeV})^2 \; .
\end{equation}
The limits from $d(Hg)$ are in this case weaker than those in (32) and
(33), since $g_p^{(s)}$ and $g_n^{(s)}$ are small.

\subsection{$\bbox{\lambda^*_{121} \lambda^\prime_{233}}$ and
$\bbox{\lambda^*_{131} \lambda^\prime_{333}}$}

The best limits come again from $d(T\ell)$ [Eq. (4)].  Using the value
(26) for $f^{(b)}_p$ and $f_n^{(b)}$ gives
\begin{equation}
\left| {\rm Im}\left( \lambda^*_{121} \lambda^\prime_{233}
\right)\right| \lesssim 4 \times 10^{-6} (m_{\tilde{\nu}_\mu}/ 100
\mbox{ GeV})^2 \; ,
\end{equation}
\begin{equation}
\left| {\rm Im} \left( \lambda^*_{131} \lambda^\prime_{333}
\right)\right| \lesssim 4 \times 10^{-6} (m_{\tilde{\nu}_\tau}/ 100
\mbox{ GeV})^2 \; .
\end{equation}

In obtaining the bounds (30) - (35) we treated the fields in the
Lagrangian (9) as if they were mass-eigenstates, which is not the case
in general \cite{38}.  While mixing is compulsory only either among
the left-handed up-type or down-type quarks and, in the light of new
evidence, among the neutrinos, the weak eigenstates are not expected
to be identical with the mass eigenstates for any of the fields
involved.  In the presence of mixing the bounds (30) - (35) hold for
the quantities ${\rm Im} \left( \lambda^*_{1j1} \lambda^\prime_{jkk}
\right)$ multiplied by a factor, which is the product of the
appropriate elements of the mixing matrices.  If the mixings are
hierarchical, this factor would not be far from unity.

The bounds (30) - (35) depend only on the constants $f_a^{(d_k)}$.
The uncertainties in the values of $f_a^{(d)}$ and $f_a^{(s)}$ come
from the experimental value of the $\sigma$-term, the values of the
quark masses, and from $SU(3)$-breaking effects.  All these are not
likely to affect the values (17) - (20) by more than a factor of 3 -
4.  The theoretical uncertainties in $f_a^{(b)}$ are, of course,
difficult to assess.

We shall consider now the constraints on the quantities ${\rm Im}
\left( \lambda^*_{1j1} \lambda^\prime_{jkk} \right)$ in (30) - (35)
from other data.  The available bounds on the individual coupling
constants $\lambda_{ijk}$ and $\lambda^\prime_{ijk}$ \cite{9} imply
upper limits on the ${\rm Im} \left( \lambda^*_{1j1}
\lambda^\prime_{jkk} \right)$ in the range $2 \times 10^{-2} -2 \times
10^{-3}$ \cite{9}.  The electric dipole moment of the electron
$(d_e)$, which is of relevance, receives contributions involving the
same products of $\lambda_{ijk}$ and $\lambda^\prime_{ijk}$ as
$C_{Sa}$ and $C_{Pa}$ through the two-loop diagrams shown in Fig. 2
\cite{39}.  These give \cite{40}
\begin{equation}
d_e \simeq e \frac{16}{3} \ \frac{\alpha}{(4\pi)^3} \
\left(\frac{m_{d_k}}{m^2_{\tilde{\nu}_j}}\right) \ \left(\ln
\frac{m^2_{d_k}}{m^2_{\tilde{\nu}_j}}\right)^2 \ {\rm Im}
\left( \lambda^*_{1j1} \lambda^\prime_{jkk} \right) \; .
\end{equation}

For $k=1$ Eq. (36) and the experimental limit $|d_e| < 4 \times
10^{-27}$ $e$cm \cite{20} imply for $m_{\tilde{\nu}_j}$ in the range
100 GeV to 1 TeV a limit on ${\rm Im} \left( \lambda^*_{1j1}
\lambda^\prime_{j11} \right)$, which is weaker than the limits (30)
and (31) by three orders of magnitude.  For $k=2$ and
$m_{\tilde{\nu}_j}$ in the same range, the limit from $d_e$ is weaker
than the limits (32) and (33) by two orders of magnitude.  However,
for $k=3$ and $m_{\tilde{\nu}_j} =$ 100 GeV we obtain
\begin{equation}
\left| {\rm Im} \left( \lambda^*_{1j1} \lambda^\prime_{j33}
\right)\right| \lesssim 6 \times 10^{-7} \;.
\end{equation}
The limit (37) (and also the limits from $d_e$ for
$m_{\tilde{\nu}_j}$ in the range $m_{\tilde{\nu}_j} = 100$ GeV to
$1$ TeV) is more stringent than (34) and (35) by an order of
magnitude.

The couplings involved in the $P$, $T$-violating e-N interactions give
rise also, through two loop diagrams \cite{39} similar to the one in
Fig. 2, to electric and chromoelectric dipole moments for the
down-type quarks ($d_{(d_k)}$ and $d^c_{(d_k)}$, respectively).  The
best limits on $d_{(d_k)}$ and $d^c_{(d_k)}$ come from the
experimental limit on the electric dipole moment of the neutron and
the $^{199}Hg$ - atom, respectively \cite{16}.  The limits implied on
the quantities $Im (\lambda^*_{1j1} \lambda^\prime_{jkk})$ are weaker
than those from $d_e$, since all the $d_{(d_k)}$ and $d^c_{(d_k)}$ are
proportional to the electron mass (rather than to the mass of $d_k$),
and also because the limits on $d_{(d)}$, $d^c_{(d)}$, $d_{(s)}$, and
$d^c_{(s)}$ are weaker than the experimental limit on $d_e$ by two or
three orders of magnitude \cite{16}, and the limits on $d_{(b)}$ and
$d^c_{(b)}$ most likely by even more.

\section{Conclusions}

In this paper we pointed out that experimental limits on $P$,
$T$-violating e-N interactions provide stringent constraints on the
$R$-parity violating interactions in the R\hspace{-.12 in}/MSSM.
Unlike in the MSSM with $R$-parity conservation, in the R\hspace{-.12
in}/MSSM $P$, $T$-violating e-N interactions arise already at the tree
level.  We found that there are two such contributions for each
down-type quark, mediated by the $\tilde{\nu}_{\mu L}$ and the
$\tilde{\nu}_{\tilde{\tau} L}$ [Eq.  (11)]. From the experimental
bounds on $P$, $T$-violating observables in atoms and molecules the
best limits on the associated coupling constant products ${\rm Im}
(\lambda^*_{1j1} \lambda^\prime_{jkk})$ $(j=2,3; \ k=1,2,3)$ come from
the limit on the electric dipole moment of the $^{205}T\ell$ atom.
Here $j$ labels the mediating sneutrino, and $k$ the participating
down-type quark.  For sneutrino masses of 100 GeV the upper limits on
${\rm Im}(\lambda^*_{1j1} \lambda^\prime_{jkk})$ are of the order of
$10^{-8}$ for $k=1$ and $k=2$, and about two orders of magnitude
weaker for $k=3$ [see Eqs. (30) - (35)].  For $k=1$ and $k=2$ these
limits are the best available bounds on these products.  For $k=3$ the
limit from $d_e$ is more stringent by an order of magnitude.

\vspace{5 mm}

{\em Note added in proof:} After this paper was submitted to Physical
Review, two papers (Refs. \cite{Go99} and \cite{Ab99}) appeared on
fermion electric dipole moments in the R\hspace{-.12 in}/MSSM. These
papers also find that in the R\hspace{-.12 in}/MSSM one-loop
contributions to the electric dipole moments from trilinear $R$-parity
violating couplings can arise only with the participation of either
neutrino Majorana mass terms, or $\tilde{\nu}_i - \tilde{\nu}^c_i$
mixing. In Ref. \cite{Go99} a limit on the coupling constants is
derived from a two-loop contribution to the electron electric dipole
moment, which is identical to the limit from the same source in our
Eq. (37).
\vspace{.5 in}

\noindent
{\bf ACKNOWLEDGMENTS}

I would like to thank S. M. Barr, G. Bhattacharyya, and R. N. Mohapatra for
useful conversations.  This work was supported by the Department of
Energy, under
contract W-7405-ENG-36.

%
%
\begin{figure}
\centering\epsfig{file=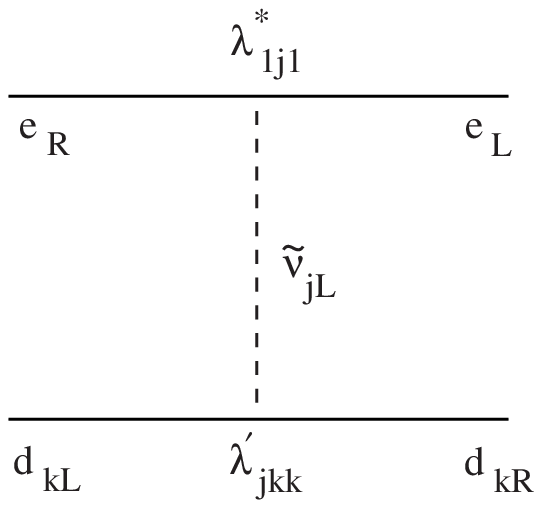,width=0.4\linewidth,clip=}
\caption[]{Diagrams contributing to P,T-violating electron-quark
interactions in the R\hspace{-.12 in}/MSSM.  Here $j = 2,3$
($\tilde{\nu}_{2L} \equiv\ \tilde{\nu}_{\tilde{\mu}L}$,
$\tilde{\nu}_{3L} \equiv \tilde{\nu}_{\tau L}$), and $k = 1,2,3$ ($d_1
\equiv d$, $d_2 \equiv s$, $d_3 \equiv b$).}
\label{fig1}
\end{figure}
\vspace{5 mm}
\begin{figure}
\centering\epsfig{file=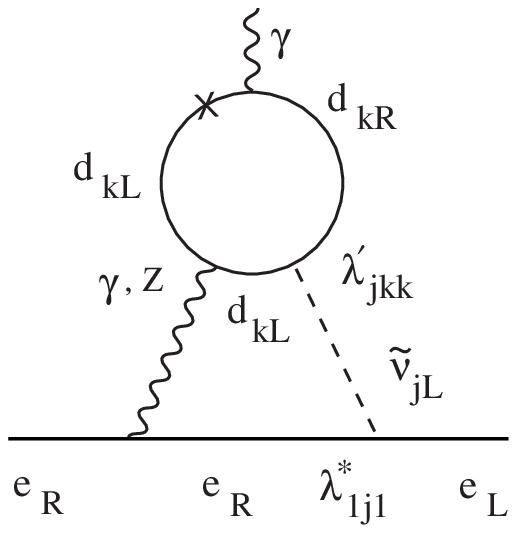,width=0.4\linewidth,clip=}
\caption[]{Two loop diagrams contributing to the electron electric
dipole moment in the R\hspace{-.12 in}/MSSM.  As in Fig. 1, $j = 2,3$
and $k = 1,2,3$.}
\label{fig2}
\end{figure}

\begin{references}
\bibitem{2} The electroweak component of the Standard Model will be
understood here to be the minimal version of the $SU(2)_L \times U(1)$
gauge theory, containing one Higgs doublet and only left-handed
neutrinos.
\end{references}
\end{document}